\documentclass[11pt,twoside]{article}

\usepackage{asp2004}
\usepackage{epsf}

\newcommand{\kms}{\ensuremath{{\rm km\,sec^{-1}}}}                   
\newcommand{\msun}{\ensuremath{\mathit{M}_{\odot}}}                  
\newcommand{\msunyr}{\ensuremath{\mathit{M}_{\odot}{\rm yr}^{-1}}}   
\newcommand{\lsun}{\ensuremath{\mathit{L}_{\odot}}}                  
\newcommand{\zsun}{\ensuremath{\mathit{Z}_{\odot}}}                  

\newcommand{\mdot}{\ensuremath{\dot{M}}}                             
\newcommand{\teff}{\ensuremath{\mathit{T}_{\rm eff}}}                
\newcommand{\vinf}{\ensuremath{v_{\infty}}}                          

\begin{document}
\title{Pre-supernova mass loss predictions for massive stars}    
\author{Jorick S. Vink$^1$, Alex de Koter$^2$ \& Rubina Kotak$^3$}   
\affil{$^1$Keele University, Astrophysics Group, Lennard-Jones Labs, ST5 5BG, Staffordshire, UK\\
       $^2$Astronomical Institute "Anton Pannekoek", University of Amsterdam, Kruislaan 403, 1098 SJ Amsterdam, The 
           Netherlands\\
       $^3$European Southern Observatory, Karl-Schwarzschild Str. 2, Garching bei M\"unchen, D-85748, Germany}    

\begin{abstract} 
Massive stars and supernovae (SNe) have a huge impact on their
environment.  Despite their importance, a comprehensive knowledge of
which massive stars produce which SNe is hitherto lacking. We use a
Monte Carlo method to predict the mass-loss rates of massive stars in
the Hertzsprung-Russell Diagram (HRD) covering all phases from the OB
main sequence, the unstable Luminous Blue Variable (LBV) stage, to the
final Wolf-Rayet (WR) phase.  Although WR produce their own metals, a
strong dependence of the mass-loss rate on the initial iron abundance
is found at sub-solar metallicities (1/10 -- 1/100 solar).  This may
present a viable mechanism to prevent the loss of angular momentum by
stellar winds, which could inhibit GRBs occurring at solar
metallicities -- providing a significant boost to the collapsar
model. Furthermore, we discuss recently reported quasi-sinusoidal
modulations in the radio lightcurves of SNe 2001ig and 2003bg.  We
show that both the sinusoidal behaviour and the recurrence timescale
of these modulations are consistent with the predicted mass-loss
behaviour of LBVs. We discuss potential ramifications for the
``Conti'' scenario for massive star evolution.
\end{abstract}


\section{Introduction}

Massive stars have a huge influence on their environments via stellar
winds and their final explosions. However, we currently do not know
with any degree of certainty, which massive stars produce which type
of supernova (SN).  The evolution of a massive star ($M$ $>$ 40 \msun)
is generally believed to be {\it driven} by mass loss, as described in
the ``Conti'' scenario: O $\rightarrow$ Luminous Blue Variable (LBV)
$\rightarrow$ Wolf-Rayet (WR) star (Maeder, this meeting), where the
WR stars include both nitrogen-rich (WN) and carbon-rich (WC) stars.

Mass loss determines the stellar mass before collapse and is thus
relevant for the type of compact remnant that is left behind
(i.e. neutron star or black hole). This process is expected to depend
on the metal content ($Z$) of the host galaxy
\cite[e.g.][]{eldridge06}. Furthermore, as WR stars are the likely
progenitors of long-duration gamma-ray bursts (GRBs), the strength of
WR winds as a function of $Z$ is especially relevant for setting the
threshold $Z$ for forming GRBs.

Given the crucial role that mass-loss plays for massive star
evolution, we have computed mass-loss rates using a Monte Carlo
method, described in \cite{abbott85}, \cite{koter97} and \cite
{vink99}.  We discuss the predictions in Sects.~2-4 in order of
decreasing temperature: WR stars $\rightarrow$ OB supergiants
$\rightarrow$ LBVs. In Sects.~5 and 6, we link our mass-loss
predictions with certain types of radio SNe.  Conclusions and outlook
are in Sect.~7.

\section{Wolf-Rayet mass-loss rates as a function of metal content}
\label{s_WR}

In recent years, it has become clear that gamma-ray bursts (GRBs) are
associated with the final explosion of a massive star, providing
enormous impetus to the collapsar model \citep{macf99}. The model
works best if the progenitor fulfils two criteria: (i) the absence of
a thick hydrogen envelope (so that the jet can emerge), and (ii) rapid
rotation of the core (so that a disk can form). This may point towards
a rapidly rotating WR star.

\begin{figure}[!th]
\plotone{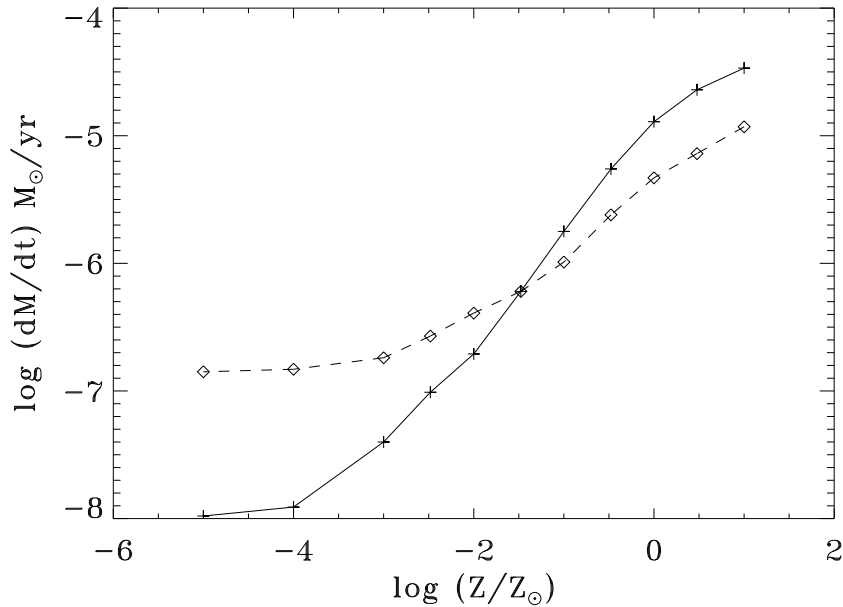} 
\caption{Mass loss versus initial $Z$ for late-type WN stars (solid
         line) and WC stars (dashed line).  Note that self-enrichment
         is accounted for, but does not enter in our expression of
         $Z$. See \cite{vink05} for details.}
\label{f_wnwc}
\end{figure}

In the so-called ``Conti'' scenario \citep{conti76}, WR stars are the
result of mass-loss during earlier evolutionary phases, while in a
complementary scenario, the removal of the hydrogen envelope may be
due to a companion.  Recently, an alternative scenario for producing a
GRB progenitor has gained popularity \citep{yoon05,woos06}: when a
star rotates rapidly, it may mix ``quasi homogeneously'', and evolve
along a track that more or less coincides with the zero age main
sequence.  A problem for producing a GRB within this scenario however,
is that the WR stars in the Galaxy possess strong winds, which may
remove the angular momentum \citep{langer98}, making it challenging,
if not impossible, to produce a GRB at Galactic $Z$.

This however might not be an issue if WR winds are weaker at low $Z$,
so the question is: ``are the winds of WR stars $Z$-dependent?'' and
if so, ``how strong is this dependence?''  The dense winds of WR stars
are likely driven by radiation pressure \citep{nugis02, graf05}, just
like their less extreme O star counterparts, which have been known to
be driven by radiation pressure since the early 1970s.  This in itself
need not necessarily imply that WR winds depend on metal content.  WR
stars produce copious amounts of metals such as carbon (in WC stars).
If, on the one hand, these self-enriched elements dominate the driving
(by their sheer number of particles), one would expect WR winds to be
independent of their initial $Z$ and the requirements of the collapsar
model may never be met. If, on the other hand, iron (Fe) is largely
responsible for the driving \citep[as in O stars;][]{vink01}, WR winds
might indeed be less efficient in low $Z$ environments.

To address the question regarding the $Z$ dependence of WR winds,
\cite{vink05} computed mass-loss rates for late-type WN and WC stars
as a function of the initial metal content (representative of the host
galaxy).  The results are shown in Fig.~\ref{f_wnwc}.  For a
discussion of the flattening in the mass-loss-$Z$ dependence for
initial metallicities below log ($Z/\zsun$) $= -2$ and potential
consequences for the first stars (Pop III), the reader is
referred to \cite{vink06}, but for the $Z$ range down to log
($Z/\zsun$) $= -2$, the mass loss is found to drop steeply, as \mdot\
$\propto$ $Z^{0.85}$, for the WN phase - where WR stars spend most of
their time. This inefficiency of WR mass loss at subsolar $Z$ may
prevent the loss of stellar angular momentum, providing a boost to the
collapsar model.

\section{Mass loss from OB stars: absolute rates and the bi-stability jump}
\label{s_ob}

\begin{figure}[!th]
\plotone{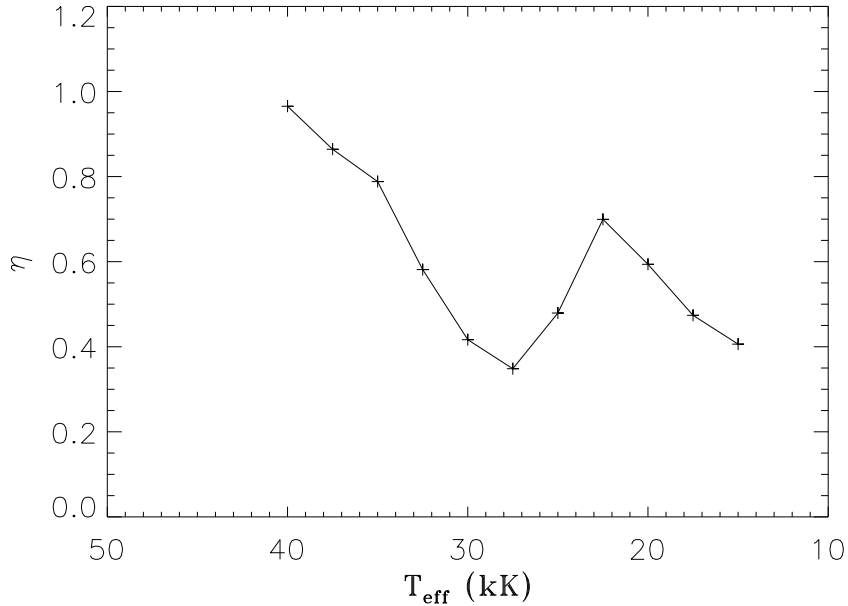} 
\caption{Wind efficiency $\eta = (\dot{M}\vinf)/({L_*/c})$ as a
         function of effective temperature. These predictions are
         taken from \cite{vink00}.  Note the presence of the
         bistability jump around 25 kK, where $\eta$ increases as Fe
         recombines to Fe\,{\sc iii}. }
\label{f_eta}
\end{figure}

We switch from a discussion of mass-loss versus $Z$ to a discussion of
mass loss versus \teff. This is best described in terms of the wind
efficiency number $\eta = (\dot{M}\vinf)/({L_*/c})$, a measure for the
momentum transfer from the photons to the ions in the
wind. Here \vinf\ is the terminal velocity of the outflow and $L_*$
the luminosity of the star.
\cite{vink00} computed wind models as a function of effective
temperature as shown in Fig.~\ref{f_eta}.  The overall behaviour is
one of decreasing $\eta$ with decreasing \teff\ due to a growing
mismatch between the wavelengths of the maximum opacity (in the UV)
and the flux (moving to longer wavelengths).  The behaviour reverses
at the ``bistability jump'' (BSJ; e.g. Lamers et al. 1995), where
$\eta$ increases by a factor of 2-3, as Fe\,{\sc iv} recombines to 
Fe\,{\sc iii} \citep{vink99}.

\begin{figure}[!th]
\plotone{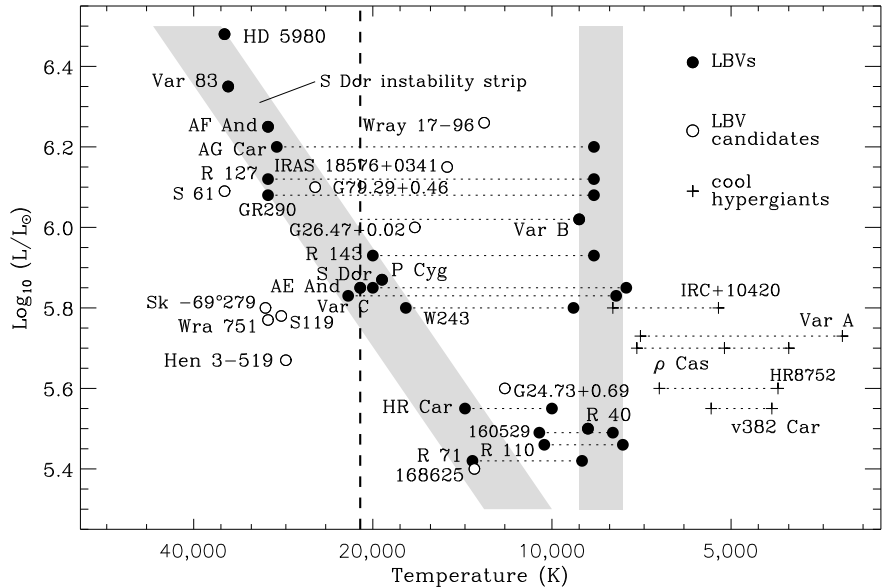} 
\caption{The LBVs in the HRD. The shaded areas represent the S~Dor
         instability strip (diagonal) and the position of the LBVs
         during outburst (vertical). The dashed vertical line at
         21\,000 K indicates the bistability jump. Figure adapted from
         \cite{smith04}.}
\label{f_smith}
\end{figure}

Recent mass-loss studies \citep{trundle05, crow06} have reconfirmed
discrepancies between empirical mass-loss rates and predictions for B
supergiants \citep{vink00}.  Discrepancies have also been reported for
O stars \citep{ful06}, and it is as yet unclear whether the reported
discrepancies for B supergiants could be due to model assumptions
(e.g. the neglect of wind clumping) or the physical reality of the
BSJ.  The most accurate way to derive \mdot\ is believed to be through
radio observations.  Intriguingly, \cite{benag07} present empirical
radio mass-loss rates as a function of effective temperature that show
a similar behaviour to the mass-loss efficiency predicted by
\cite{vink00}. This may well be the first evidence of the presence of
a mass-loss BSJ at the boundary between O and B supergiants.  The
relevance for stellar evolution is that when massive stars evolve to
lower \teff\ after the O star main sequence phase, they are expected
to cross the BSJ. Interestingly, LBVs brighter than log ($L/\lsun$) $=
5.8$ (see Fig.~\ref{f_smith}) are expected to encounter it
continuously - on timescales of their photometric variability, which
we discuss in the next section.

\section{Mass loss from Luminous Blue Variables}
\label{s_lbv}

\begin{figure}[!th]
\plotone{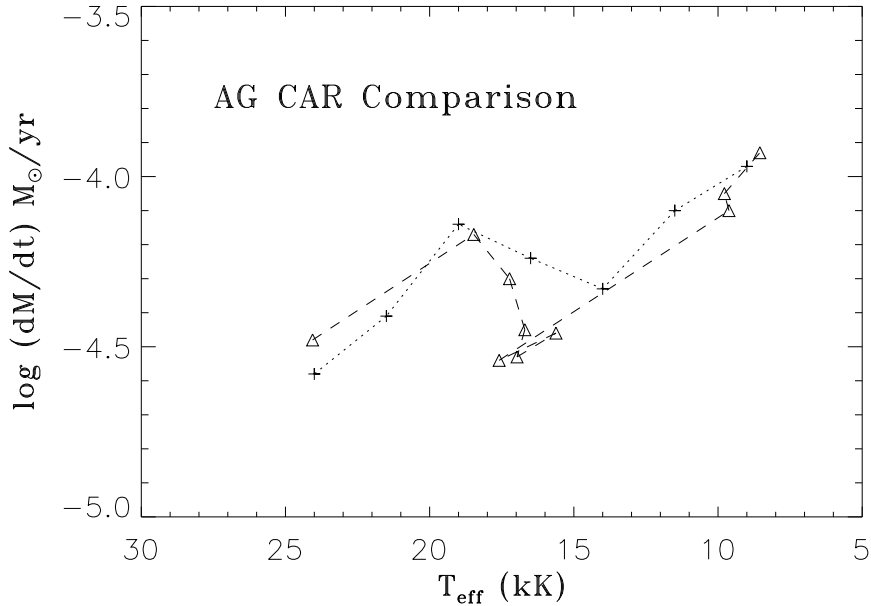} 
\caption{Predicted (dotted line) and empirical (dashed line) mass-loss
         rates versus \teff\ for the LBV AG~Car. Note that both the
         qualitative behaviour and the amplitude of the mass-loss
         variations are well reproduced. See \cite{vink02} for
         details.}
\label{f_ag}
\end{figure}

LBVs are unstable massive stars in the upper part of the HRD
\cite[e.g.][]{hd94}.  As can be seen in Fig.~\ref{f_smith}, the
classical LBVs, like AG Car, are anticipated to cross the BSJ at
$\sim$ 21\,000 K. One of the defining characteristics for LBVs is their
S~Doradus (SD) variation of $\sim$1 -- 2 mag on timescales of years
(short SD phases) to decades (long SD phases)
\citep{vgenderen01}. \citet{vink02} computed LBV mass-loss rates as a
function of \teff - shown in Fig.~\ref{f_ag}. Overplotted are the
empirical mass-loss rates for AG Car \citep{stahl01}, which vary on
the timescales of the photometric variability.  Although the agreement
is not perfect \cite[see][for a discussion]{vink02}, the amplitude of
the predicted variability fits the observations well, and most
importantly the overall behaviour appears to be very similar, and may
indeed be explained in terms of the physics of the BSJ.  This
bi-stable behaviour in an individual stellar wind \citep{pp90} causes
the star to flip back and forth between two states: that of a low mass
loss, high-velocity wind, to a high mass-loss, low velocity wind.  The
wind density ($\propto \mdot/\vinf$) would therefore be expected to
change by a factor of $\sim$2~$\times$~$\sim$2, i.e. $\sim$4 on the
timescale of the SD variations. In the absence of any other material
around the star, this would result in a pattern of concentric shells
of varying density.

\section{Radio supernovae and progenitor mass loss}
\label{s_radio}

\begin{figure}[!th]
\plotone{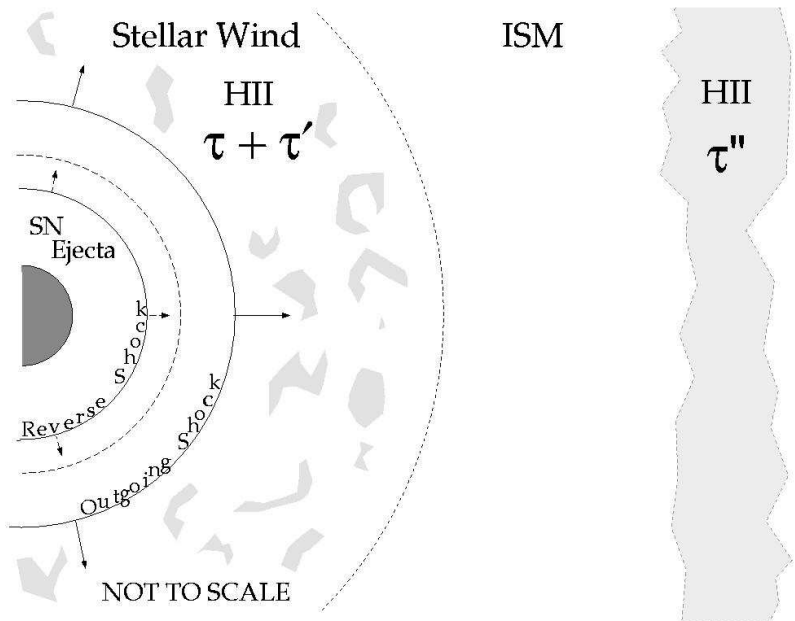} 
\caption{Model for SN ejecta interaction with the progenitor's wind.
         The radio emission arises at the interface between the
         outgoing shock and the most recent stellar wind.  The various
         optical depth ($\tau$, $\tau'$, $\tau''$) contributions are
         respectively from a smooth wind, a clumped wind and a
         potential intervening H\,{\sc ii} region. Taken from
         \cite{weiler02}.}
\label{f_sketch}
\end{figure}

Radio SNe (RSNe) lightcurves and the model for SN interaction with the
surrounding circumstellar material has been reviewed by
\cite{weiler86}, and is shown schematically in
Fig.~\ref{f_sketch}. The radio emission is due to non-thermal
electrons, while the absorption may be due to both synchrotron self
absorption as well as free-free absorption. Examples of the rise,
peak, and power-law decline of radio lightcurves are shown in
Fig.~\ref{f_soder}.  (The episodic bumps at late time are discussed in
Sect.~\ref{s_quasi})

The model constrains the wind density and thus the ratio of \mdot\ to
the terminal wind velocity: $\rho \propto \mdot/\vinf
r^2$. Assuming $\vinf$, \cite{weiler02} list \mdot\ values in the
range $10^{-6}$--$10^{-4}$ \msunyr. Fortunately, these values agree
with mass-loss predictions, but are broadly representative for massive
stars over almost all post-main sequence evolutionary phases, making
it hard to infer the progenitor from radio lightcurves alone, unless
these lightcurves betray their progenitor in some other way.

\section{Quasi-periodic oscillations in radio SNe lightcurves}
\label{s_quasi}

\begin{figure}[!th]
\plotone{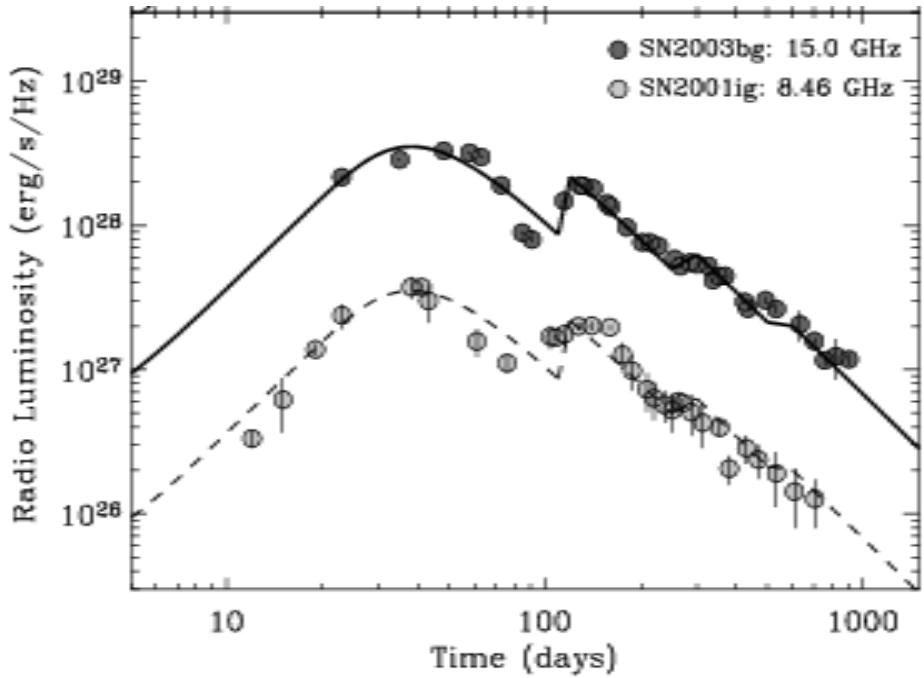} 
\caption{Radio luminosity versus time for two strikingly similar
         recent SNe: 2001ig and 2003bg.  Note the quasi-sinusoidal
         modulations during the power-law decline phase. Taken from
         \cite{soderberg05}.}
\label{f_soder}
\end{figure}

A number of recent RSNe have shown sinusoidal modulations in their
radio lightcurves, in particular SN\,2001ig \citep{ryder04} and
SN\,2003bg \citep{soderberg05} are strikingly similar in terms of both
amplitude and variability timescale (see Fig.~\ref{f_soder}).
The recurrence timescale $t$ of the bumps is $\sim$ 150 days.  Using
Eq.~(13) from \citet{weiler86}:

\begin{equation}
\Delta P~=~\frac{R_{\rm shell}}{v_{\rm wind}}~=~\frac{v_{\rm ejecta}~t_{\rm i}}{v_{\rm wind}~m} \left(\frac{t}{t_{\rm i}}\right)^m
\label{eq:period}
\end{equation}
where $m$ is the deceleration parameter (here $m$ = 0.85) and $t_{\rm
i}$ is the time of measurement of the ejecta velocity relative to the
moment of the explosion.  Assuming $v_{\rm wind}$ = 10--20\,\kms,
typical wind velocities for red (super)giants, \citet{ryder04} found a
period $P$ between successive mass-loss phases that was too long for
red (super)giant pulsations (100s of days), but too short for thermal
pulses (10$^2$--10$^3$ years). They therefore invoked an edge-on,
eccentric binary scenario involving a WR-star and a massive companion.
One of the main differences between LBV and red giant winds is that
LBV winds are about 10 times faster. If the progenitor of SN\,2001ig
were an LBV, the expected period between successive mass-loss episodes
would be $\Delta P \sim$\,25\,yr (for an assumed $v_{\rm
wind}$~=~200\,\kms), consistent with the long SD phase
\citep{kotak06}.

\cite{soderberg05} infer density enhancements of a factor of $\sim$2
during the deviations from pure power-law evolution.  They consider a
range of options that might account for the modulations, but they
favour a single-star progenitor model of a WR star that underwent
episodes of intensified mass loss. However, they do not specify the
physical mechanism that gives rise to these periods of enhanced mass
loss. Our SD mechanism for LBVs may alleviate this shortcoming.

\section{Discussion: do LBVs explode?}
\label{s_concl}

Are LBVs viable SNe progenitors?  It may be relevant that both
SNe\,2001ig and 2003bg are ``transitional'' objects.  SN\,2001ig was
initially classified as type II (showing H lines) but metamorphosed
into a type Ib/c object (no H lines, weak He lines) about 9 months
later.  This suggests that it has lost most of its H-rich
envelope. SN\,2003bg however was first classified as a type Ic, but
within a month the spectrum evolved into a type II SN. This
transitional behaviour hints at the fact that their progenitors are
intermediate evolutionary objects: H-rich compared to OB/red
(super)giants, but H-poor compared to WR stars.  LBVs are likely
candidates.

During this meeting, there was discussion about clumping in O star
winds. The value for the clumping factor in O star winds is very
much an open issue. If these factors would be much larger than
two, this would have severe implications for massive star evolution.
One consequence might be that giant LBV eruptions ($\eta$ Car type
eruptions, not the typifying SD variations) dominate the integrated
mass loss during evolution \citep{smith06}.
An alternative scenario could be that post-main sequence stars do not
become WR stars, but explode early -- during their LBV phase.

Here, we have presented indications that at least those SNe that show
quasi-periodic modulations in their radio lightcurves might have LBV
progenitors \citep{kotak06}.  It has also been speculated that LBVs
may be the generic progenitors of type IIn SNe \citep{gal07}, however
the type IIn phenomenon (arising from SN ejecta expanding into dense
circumstellar matter) may be relevant to both core-collapse and
thermonuclear SNe \citep{kotak04}, and although it may be reasonable
to expect that some type IIn SNe have LBV progenitors, there remains a
lot of work to be done to prove this.
Nonetheless, it appears that the ``standard scenario'' for massive
star evolution may need revision. Future mass-loss predictions will
certainly play a major role in adjusting even our most basic knowledge
of massive star evolution.

\acknowledgements 
Henny, thanks for sharing your insight in physics and unlimited enthusiasm.


\end{document}